*Research Article*

# The Response of Farmer Welfares Amidst Food Prices Shock and Inflation in the Province of East Java

[1]**Moh. Hairus Zaman**, [2]**Diah Wahyuningsih**, [3]**Ris Yuwono Yudo Nugroho**
[1,2,3]*Department Master of Economics, University of Trunojoyo Madura, Indonesia.*

***Abstract:*** *Price uncertainty in food commodities can create uncertainty for farmers and potentially negatively impact the level of farmer household well-being. On the other hand, the agriculture sector in the province of East Java has greatly contributed to East Java's economy. This paper analyses the response of farmer welfare through farmer exchange values amidst fluctuation shock of food needed prices and inflation level in the east java province. The research method of this paper employs the impulse response function of the Bayesian Vector Autoregressive (BVAR) model by using time series secondary data from May 2017 until December 2023. This paper finds that the shock that happens to aggregate food prices can increase farmer exchange values even though the shock to the inflation level has reduced farmer exchange values and increased aggregate food prices.*

***Keywords:*** *Agriculture Sector, Aggregate Food Price, Inflation, Farmer Welfare.*

## I. INTRODUCTION

The province of East Java is one of the provinces in Indonesia that has a strategic role in supporting the stability of prices and food security. The province of East Java has a wide region, with most residents being farmers. Based on the result of the agriculture survey of the Central Bureau of Statistics (BPS) in the year 2023, around 5.372.003 households, the total productivity of farmer households who are workers in agriculture sectors is approximately 971.102 workers. Those facts show that the province of East Java has the highest productivity of workers of the age of employment in the agriculture sector compared with other provinces in Indonesia (Central Bureau of Statistics (BPS), 2023). It reflects that the many residents in the province of east Java hang up their livelihood for agriculture sector activities.

The farmers of East Java province have an important role in maintaining national and regional food security. Nevertheless, both the food price shock and inflation level cannot be controlled, leading to uncertainty and the potential to reduce the agriculture productivity level and farmers' welfare level. On the other side, the farmer is met with many more internal and external challenges, such as state conditions, extreme weather, pests, plant diseases, changes in trade policies and global economic circumstances. All of that has an impact up down on food prices and inflation levels (Ivanic & Martin, 2008; Mankiw, 2020).

Food price shocks have impacted not only consumers but also farmers as producers. On the other side, food price shock will increase the farmer's income. However, it leads to uncertainty for farmers. Food prices tend to fluctuate, and in odd moments, farmers decide to plant certain crops because of worries about financial risks that would be invested (Barrett et al. 2010). Food price shock can be described through price elasticity, namely how much of the quantities are demanded or supplied. Increasing prices have low demand elasticity for basic goods, which will increase farmers' income levels. However, when high elasticity would lead to the demand side deeply reducing and restricting against the increasing farmer income (Nicholson, 2005).

Food price shocks are able to encourage the increase of inflation level and lead to a spillover effect on increased other goods prices because of increasing production costs (Blanchard & Johnson, 2017). According to the Food and Agriculture Organization (2011) report, food price shock is due to the increasing inflation level, especially in developing countries. Because the food needed is high spending of households (FAO, 2011), the households that have an under-middle income will be very impacted by the food prices. Therefore, it led to reduced purchasing power consumption that impacted on happening the economic uncertainty (Dercon, 2005).

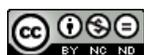

**The level of farmer exchange value (NTP)**

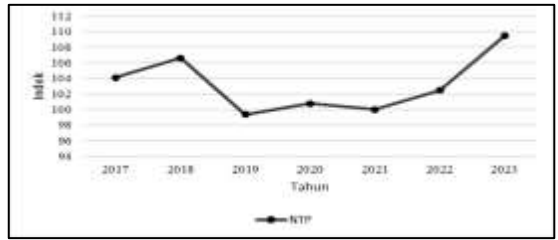

**Fig 1. The level of farmer exchange value (NTP)**

In the last seven years, from 2017 until 2023, the level of farmer's exchange value in east Java has shown a trend of shapes like curve-U. Figure 1 shows the level of farmers' exchange value from 2017 until 2023 in East Java. Figure 1 shows that the level of farmer's exchange value has increased during the period 2017 until 2018, approximately 104 to 106. It subsequently happened deeply reduced during the period 2018 until 2021, around 100 index in last 2021. In the further year circumstance, the level of farmer's exchange value has sharply increased during the period between 2021 and 2023. However, it does not mean that a high level of farmer's exchange value nominally does not automatically describe the qualities of farmer welfare condition is real empirically. On the other side, it happens amidst the shock of food prices and inflation levels (Krugman & Obstfeld, 2009). Furthermore, all of those need to be examined even further to understand the relationship between food, inflation, and farmer exchange value and obtain an accurate description of the farmer's welfare.

According to Rata et al. (2012) and Nasution et al. (2022), the farmer's income is to buy goods and services for consumption, with goods and services resulting from farmers for sale defined as part of the instrument of farmer exchange value. Therefore, high farmer exchange values have been described as farmer circumstances and well-being. When the farmer's income is on needed consumption, it is usually followed by the high farmer exchange value, and vice versa (Syifa Aulia et al., 2021). According to Saridewi (2021) explains that inflation has become one of the factors that influence farmer exchange values.

On the other side, Umaroh et al. (2019) found that increasing food prices impacted reducing farmer well-being. According to Pratomo et al. (2023), there is an anomaly associated between food prices and farmer welfare conditions. The high food prices in the market are not fully profitable for farmers. On the other side, Yuliana et al. (2019) reveal that the increase in food prices in Indonesia has impacted reducing welfare, both non-farmer households and farmer households. At the same time, different findings in other countries such as China and Nigeria, the increase in food prices has a diverse impact on farmer welfare circumstances (Adekunle et al., 2020; Jiao-hua & Chang-jian, 2013; Yu, 2018).

Consequently, this paper has an obvious research gap compared to previous research. This paper attempts to employ the *Bayesian Impulse Response Function* (BIRF) approach by utilizing the *Bayesian Vector Autoregressive* (BVAR*) model. According to* Bańbura et al. (2010), the BVAR model is able to predict better, especially in the short and middle term, as well as more robust compared with the conventional VAR model. On the other side, this paper not only focuses on the response of farmer welfare to food price shock happens, but also considering to examine the response of farmer welfare to inflation shock in the province of east Java.

Based on the explained above, this paper is conducted by several purposing as follows. Firstly, to determine aggregate food prices variable based on food community prices as well as ensuring data used for all variables has done stationarity in level. Secondly, analyzing the response of farmer welfare circumstances through the farmer exchange values indicator at the time of the aggregate food prices and inflation shock. Thirdly, it will analyze the response of inflation to the aggregate shock in food prices.

## II. LITERATURE REVIEW

In the basic needed context, excessive demand is able to produce high basic needed prices or increase inflation in the market. The instability of food commodity prices is due to a lot of factors such as government policy, goods qualities, disasters, and important moments (Vermila, 2016). Food price shock is potentiated on food security and fragile people communities such as traditional farmers and poor people who have limited access to the market (Kalkuhl et al., 2016).

*A) The Change in Food Prices*

Theoretically, when there is no change of structure on goods and services, the change of prices is signed by directly altering the whole goods and services level. According to Suwardjono (2005) explains that the change in prices reflects up or down from currency value, or that is called purchasing power. Other factors often occur; there is an imbalance between the demand and supply of general goods and services.

Goods and services prices in a competitive market are determined by demand and supply. Demand behavior is one of the dominant behavior in macroeconomic practicals. Therefore, demand discourse in determining price through the demand perspective will always be the core discourse in economics. On the other hand, the supply side is a certain number of goods and services average to sale at another probability price in a certain short time, ceteris paribus. The supply side shows maximalized total willingness for sale at different price levels or certain minimum prices that encourage sellers to offer a number of certain goods (Todaro & Smith, 2020).

On the other side, price shock reflects a perception that price fluctuates in the long term or is a stable enough price trend (Hull & Basu, 2016) in the short term. This fluctuation can refer to prices daily, weekly and monthly. Commodity prices are lower or higher; it is often associated with crisis circumstances that lead to challenges for producers, consumers, and policymakers. Volatility design touches on the price shock notion in two different ways, namely the historical approach and the prediction approach.

According to Huchet-Bourdon (2011), price shock is a natural phenomenon, and food price uncertainty is a risk for developing countries. Because when food prices increase, it will increase the inflation level and reduce economic growth. However, those circumstances have a potential impact on households, especially poor people. Price shock and outcome are interconnected with each other, and the response supply side against price shock depends on the producer's ability to take a risk (Newbery & Stiglitz, 1981; Sandmo, 1971; Subervie, 2008). Especially farmers who are afraid to take a risk and tend to choose harder work to increase their income. Farmers need protection from bad weather, but farmers are scared to take risks and tend to reduce production levels.

*B) Farmer exchange values*

Central Bureau Statistical (BPS) officially publishes the farmer exchange values (NTP) as an indicator to gauge farmer purchasing power and farmer well-being. According to Syekh (2013), farmer exchange values have existed since the 1980s, and they are indicators of farmer well-being to see the performance of farmer income ability. The farmer exchange value is a comparison price index received by the farmer (It) with the price index bought by the farmer (Ib) by using percentages unit. (It) is producer growth of price index as a result of farmer production. It shows the fluctuation of goods produced by farmers. On the other hand, it is also able to be used as farmer income balance data in agriculture sectors. Whereas (Ib) is the price index which is needed by farmers consisting of farmer household consumer needs and requirements of the agriculture production process. Besides that, this channel is able to show the fluctuation of goods prices that are often consumed and support farmer production processes. The equation of this farmer exchange values can be predicted as follows:

$$\text{NTP} = \left(\frac{It}{1b}\right) x\ 100\% \qquad (1)$$

Where NTP is the farmer exchange values, *It* is the price index received by farmers, and *Ib* is the price index bought by farmers. As a result, equation accounting number 1 (one) means that higher farmer exchange values show better farmer purchasing power on product consumption as well as showing the farmers have a welfare relative category, and vice versa. (Badan Pusat Statistik, 2024).

Several previous types of research have been done to describe the associated change in food price circumstances and inflation on farmer well-being condition as well as food price contribution on inflation level. Research associated with food price contribution and inflation was conducted by Mpofu (2017), who attended to analyze the relationship between macroeconomic variables and food price inflation and nonfood price inflation and compared with circumstances in other countries in South Africa. It utilizes multiple regression methods and time series data used in the beginning 2010 to 2016. The results of the research show that the determinant of the relation level between variables used has implications for food security and all resident welfare aspects in Zimbabwe. Spending on nonfood is a significant component of the household budget, and increasing prices often produces reduced purchasing power and household security. Increasing food prices in Zimbabwe has a low relation to economic macro variables such as money supply, dollar currency exchange, and inflation.

Peersman (2022) attempted to examine the impact of the change in international commodity food prices on inflation dynamics in the European region by employing the *structural Vector Aggressive Model* (SVAR) that is identified with external instruments such as global agriculture shock. The result shows that food commodity prices shock as exogen variables have a strong impact on consumer prices by inflation volatilities around 25% to 30%. On the other side, huge increases in international food prices have significantly contributed to the puzzle of problems reduced and increasing inflation after a huge recession era. Without destroying the global food market, inflation in the European region has been lower in the period 2009 to 2021 and higher in 2014 to 2015, respectively. Transition mechanism analysis shows international food price shock has an impact on retail prices through the food production chain but also triggers an indirect impact through inflation expectation and European currency depreciation.

Yuliana et al. (2019) have conducted research in Indonesia associated food price issues and welfare based on clarification of different groups, such as residents in urban and rural, poverty status, main income source, and poverty status of non-farmer and farmer households. By employing several analyzing models like Compensating Variation and Linear Approximation Almost Ideal Demand System (LA/AIDS) model, the result of the research shows that when happening increased food prices are low, the whole group experiences reduced welfare levels as well, and the agricultural household has higher reduced than non-farmer households. As a result, this research reveals that rice food commodity has a high contribution to reducing welfare, especially for farmer household groups.

On the other side, research undertaken by Adekunle et al. (2020) tries to analyze the multiple roles of farmer households as producers and consumers in food commodities to find the impact of price changes on farmer well-being in Nigeria country, by utilizing *Quadratic Almost Ideal Demand System* (QUAIDS) dan *Compensating Variation* (CV) methods by using two types of data namely household consumption data and grosser commodity prices in 36 countries in data panel term during period 2007 to 2016. The finding explains that consumption patterns and production, directly and indirectly, impact farmer household welfare in Nigeria. Furthermore, 79% of farmer households are consumers of food commodities. Therefore. When food prices are high, farmer households experience reducing or indirect impact. 21% of food producers from rows of agriculture obtain increasing welfare while increasing food prices.

Specifically, research was already done by Umaroh et al. (2019) in the province of East Java to analyze the impact of the change in food prices on general welfare, including farmer welfare in East Java. That research employs microdata obtained from the *Indonesian Family Life Survey* (IFLS) between the period 2000. 2007 and 2014, using the *Quadratic Almost Ideal Demand System* (QUAIDS) and *Compensating Variation* (CV) methods. The result of the research shows that the impact of increasing food prices has negatively influenced general welfare. Reducing welfare is lower in the number of poor households and farmer households in urban than poor households and non-farmer households in urban. Because farmer households have economic activities in agriculture sectors when the food prices increase, farmer household welfare obtains a good incentive to sell yields; when happening the increased food prices increase as well, it is able to choose to consume their own yields. It is subsequently able to be minimalize the impact of the increase in food prices.

Shittu et al. (2014) also conducted research in regard to the transmission of monetary policy variables to the change of food prices on farmer welfare in Nigeria. By accounting for the cointegration relation between consumer price and policy variables, vectors are used as exogen variables, and demand elasticity is to be accounted for from the equation system of household demand. Those asses welfare from policy impact leads to the increasing of food price on agriculture household. The result of research reveals that the management of government from interest rate and money supply, as well as withdrawal of subsidies from crude oil products, lead to an increase in food prices. At the same time, the increase in food prices is due to a compensation policy from the government for farmers. However, it causes most farmer households to end up losing.

On the other side, Yu (2018) has conducted research in China, trying to review the background of theory, methodology, and empirical application of the Engel curve to be used to examine the change in farmer welfare and food demand after economic reformation in 1978 and comparing by income statistic and food consumption. The methodology used is the Engel curve, which compares income growth and food consumption to research the increase in farmer welfare in rural China by using data from 40 years. It found that in the beginning period from 1978 until 1988, farmer welfare increased due to the economic reformation gift in 1978 as well as the increasing purchase of government agriculture products. In the second period, from 1989 to 1995, farmer welfare was reduced, causing the ending of institution reformation gift. The increasing of food prices, relatively high inflation, and unstable political conditions. In the third period, after 1995, farmers' welfare came back to increase due to multiple price systems being erased, the transition from economic planning to market economics was erased, and the government conducted policy protection for agriculture and started to provide huge agriculture subsidies.

Previous research done by Jiao-hua and Chang-jian (2013) attempted to examine the impact of agriculture commodity price shock on farmer income in China. Research data was used from 1979 until 2010 by employing the VAR model. The findings of the research conclude that fluctuation of agricultural commodity prices significantly influences farmer incomes. However, the increase in farmers happens in the short term, and farmer incomes experience a reduction in the long term.

### III. RESULTS AND DISCUSSION
*A) Identifying Common Factors, Such as Aggregate Food Prices*

Figure 2 reports aggregate food prices and whole food prices; both describe comparing aggregate food prices as factor common variable with each food price such as rice price, chicken meat price, beef price, egg price, red onion, garlic, cayenne pepper price, cooking oil price, and sugar price in the province of east Java. Comparing is looked at during the period of May 2017 until December 2023. Every move of price fluctuation describes price shock based on unit percentage in all food prices.

Based on Figure 2, it can be explained that rice prices during the period of May 2017 to December 2023 experienced price shock, with the lowest reducing values happening in February 2019 and experiencing the highest growth increase for rice prices by approximately 11% on March 2019 by an average of rice price growth around 0,48%. Chicken meal price shock happened from 2017 until 2023 experienced the highest chicken meal price increase in June 2020, around 34,3%, reducing the lowest price growth that happened in September 2018, by an average of chicken meal price from May 2017 until December 2023, fluctuating about 0,79%.

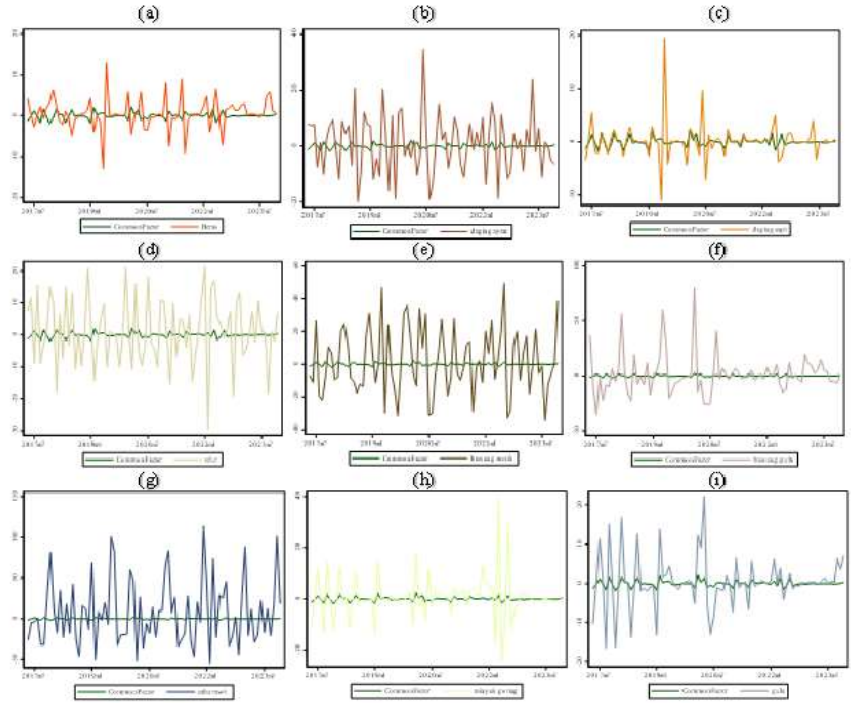

Note: (a) Is comparing common factor and rice price, (b) is comparing between common factor and chicken meal price, (c) is comparing between common factor and beef price, (d) is comparing between common factor and egg price, (e) is comparing between common factor and red union, (f) is comparing between common factor and garlic price, (g) is comparing between common factor and cayenne pepper price, (h) is comparing between common factor and cooking oil price, and (i) is comparing between common factor and sugar price.

**Figure 2. Aggregate food price in all of food commodity price**

Whereas the growth of beef price experienced an increase to the highest price shock in June 2019, approximately 19,39%, the lowest growth of beef price was around -11,03% in Mei, 2019. The average growth of beef prices is around 0,15%. Furthermore, egg price shock increased the highest price growth in January 2022 and the lowest price growth in February 2022, approximately 19,39% and -29,53%, respectively. The average of egg prices from 2017 to 2023 was around 1,12%.

On the other side, the growth of cayenne pepper rice and cooking oil have increased by approximately 113,09% and 40%, respectively. The highest growth of cayenne pepper prices happened in December 2023, and the highest growth of cooking oil shock happened in April 2022. On the other hand, the growth of cayenne pepper prices and cooking oil prices lowest reduced in February 2022 at around -53,92% and in May 2022 at around -24,67%, respectively. On average, the growth of cayenne pepper prices and cooking oil prices was approximately 7,54% and 0,74%, respectively.

The growth of sugar price shock increased by around 22,22% in April 2020, with the lowest reduction of about 16,51% in September 2017. The average growth of sugar prices is around 0,459%. On the other side, the average growth of aggregate food price is around 0.00000000128%, with the highest and lowest price growth of approximately 2,07% and -2,02%, respectively. According to Ismaya & Anugrah (2018), the increase in aggregate food prices can happen because of several factors such as food production, the result of the agriculture sector, infrastructure, food imported, credit of the agriculture sector, consumption demand/supply money, and important days.

*B) Stationery test*

The result of The Augmented Dicky Fuller (ADF) test identifies that all of the variables used in this paper stationery in level. Empirical been analyzed previously, and this paper employs a unit root test to examine the stationarity of all variables that are used in the VAR model to eliminate the influencing of random law differentiator in non-stationery series (Apergis et al., 2020 dan Diks & Panchenko, 2006). The result of the stationery test can reference the probability values of each variable in Table 1, where variables data can be mentioned stationery when p-value < 5%.

**Tabel 1: Uji Unit Root**

| Variable | ADF-Statistic | 1% | 5% | 10% | *p-value* |
|---|---|---|---|---|---|
| NTP | -5.539 | -3.539 | -2.907 | -2.588 | 0.000*** |
| INF | -6.405 | -3.539 | -2.907 | -2.588 | 0.000*** |
| HPA | -11.444 | -3.539 | -2.907 | -2.588 | 0.000*** |

*Note*: NTP is the farmer exchange values, INF is the inflation rate, and HPA is aggregate food prices

*C) Analyzing the Result of Impulse Response*

Based on the result of common factor extraction to food price and the result of unit root test in all variables, this paper would like to further explore three dimensions of the impulse response function: first, the response of farmer exchange values to aggregate food prices and inflation shock. Second, the response of inflation level to food price shock.

Based on figure 3 shows that when aggregate food price shock happens, the inflation level responds passively and is stagnant, located around the equivalence line until the last period of time. Based on figure 3 shows that aggregate food price shock has an influence and is not significant enough on the increasing inflation level. That result confirms that it is suitable to previous research done by Mpofu (2017), who found empirical evidence that thtype of nonfood prices more strongly triggers the increase in inflation in Zimbabwe than the increase in food prices. According to Huchet-Bourdon (2011) argued that in developing countries usually occur unstable food prices which will influence increasing inflation. Nevertheless, research conducted by Peersman (2022) in the European region explains that the increase in food prices at the retail level happens because it is caused by the international food price shock that indirectly impacts the increase of inflation through currency depression and increasing wages.

On the other side, the findings in this research confirm suitability with research conducted by Jiao-hua and Chang-jian (2013), who found that the change in food commodity prices has a significant influence on farmer welfare. Based on figure 3 explains that farmer exchange values respond positively directly to aggregate food price shock in the beginning period until 2nd month, around 15%, while subsequently experiencing a trend that reduces during 3rd month until the last period, approximately 3%. Although the trend has reduced, the response of farmer exchange values does not touch the convergence line until the last period. This means that the increase in aggregate food prices can directly increase farmers' exchange values. On the other hand, the majority of residence provisions in East Java are for employers in the agriculture sector. So, aggregate food price shock can influence farmer welfare conditions in East Java. However, those empirical findings are opposite to previous research done by Adekunle et al. (2020) and Chigozirim et al. (2021) in Nigeria, that the dynamic of food price shock alters household consumption that negatively impacts farmer welfare level and general household in the short term. The increase in food prices in Nigeria was not much enjoyed by direct farmers but was much more profitable for producers or businessmen who use ingredients produced by farmers. Those findings also confirm differences from research done by Wossen et al. (2018) in Ethiopia and Ghana, which reveals that farmer household incomes have significantly reduced due to food price shocks that can potentially influence food security.

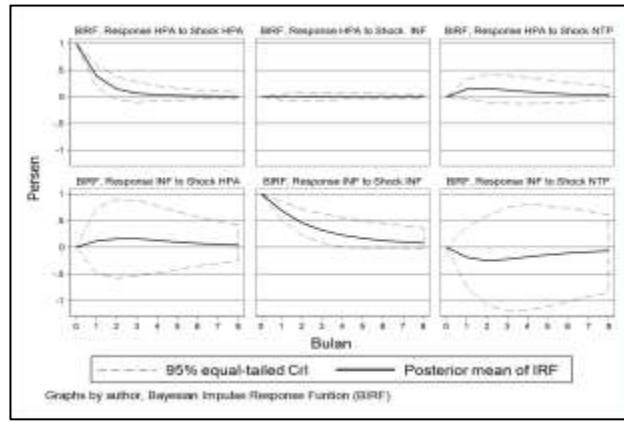

**Figure 3. Response of farmer exchange values and inflation to aggregate food prices shock**

The response of inflation is happening shock to itself; direct response with the trend has reduced since the beginning period, around 100%, and directly moving toward the convergence line until the last period, around 7%. Different from a shock on food princes, when happens in inflation, the farmer exchange values responses in negative since in the beginning period to 3rd month with the lowest response approximately -25%, while regularly increased. However, it is not able to touch the convergence line until the last period, around 6,8%. It implies that the increase in goods and services prices causes reducing in the abilities of prices that farmers on diverse farmer buy needed. Hence, it directly impacts reducing on farmer welfare when happening shock to inflation. On the other hand, inflation has a shock, responding by positively aggregating food prices from the beginning period until the 3rd month at around 16%, then continuing to experience a reduction in the last period at around 4%. However, the response of aggregate food prices to inflation shock is not able to experience convergence until the last period. This implies that when this happens, an inflation shock will not only reduce farmer exchange values but will also lead to an increase in aggregate food prices. According to Stiglitz & Rosengard (2015), the increasing inflation level can destroy farmer welfare due to the increasing living costs, reducing real purchasing power. However, farmers still benefited because it has multiple roles as producers and consumers. Hence, when an inflation level shock happens, farmers can still enjoy the result of their agriculture directly to fulfil the basic needs of farmer households (Shittu et al., 2014; Umaroh et al., 2019; Yuliana et al., 2019).

Those results also confirm that it is different from the result of empirical findings previously researched in Indonesia, which state that the increase in prices at the consumer level influences farmer welfare (Syifa Aulia et al., 2021). Nevertheless, Y X (2018), who has researched rural areas of China, revealed that the increase in food prices and inflation reduce farmer welfare in China. Farmer welfare in rural areas started regularly increasing when the government enforced a policy to save the agricultural sector and started to provide subsidies for farmers.

## IV. CONCLUSION

Price shocks that happen in aggregate food prices positively impact farmer exchange values in the province of east Java while shocks in inflation level show different results. The aggregate food price shock cannot trigger an increase in the inflation level. It is enabled because the increase in inflation is triggered by the increase in nonfood prices. The increase in aggregate food prices is able to provide a positive impact on the increase of farmer income in the short term. However, the increase in income for farmers is still fragile to happen reducing. Particularly, when the inflation level has increased, it means that farmers' incomes cannot reach the increase of goods and services that caused the low. On the other side, inflation level shock not only reduces farmer exchange values but also leads to the increasing of aggregate food prices.